# SPEED: the Segmented Pupil Experiment for Exoplanet Detection

P. Martinez[*a], O. Preis[a], C. Gouvret[a], J. Dejongue[a], J-B. Daban[a], A. Spang[a], F. Martinache[a], M. Beaulieu[a], P. Janin-Potiron[a], L. Abe[a], Y. Fantei-Caujolle[a], D. Mattei[a], and S. Ottogalli[a]

[a] Laboratoire Lagrange, UMR7293, Université de Nice Sophia-Antipolis, CNRS, Observatoire de la Côte d'Azur, Bd. de l'Observatoire, 06304 Nice, France

## ABSTRACT

Searching for nearby exoplanets with direct imaging is one of the major scientific drivers for both space and ground-based programs. While the second generation of dedicated high-contrast instruments on 8-m class telescopes is about to greatly expand the sample of directly imaged planets, exploring the planetary parameter space to hitherto-unseen regions ideally down to Terrestrial planets is a major technological challenge for the forthcoming decades. This requires increasing spatial resolution and significantly improving high contrast imaging capabilities at close angular separations. Segmented telescopes offer a practical path toward dramatically enlarging telescope diameter from the ground (ELTs), or achieving optimal diameter in space. However, translating current technological advances in the domain of high-contrast imaging for monolithic apertures to the case of segmented apertures is far from trivial.

SPEED – the segmented pupil experiment for exoplanet detection – is a new instrumental facility in development at the Lagrange laboratory for enabling strategies and technologies for high-contrast instrumentation with segmented telescopes. SPEED combines wavefront control including precision segment phasing architectures, wavefront shaping using two sequential high order deformable mirrors for both phase and amplitude control, and advanced coronagraphy struggled to very close angular separations (PIAACMC). SPEED represents significant investments and technology developments towards the ELT area and future spatial missions, and will offer an ideal cocoon to pave the road of technological progress in both phasing and high-contrast domains with complex/irregular apertures. In this paper, we describe the overall design and philosophy of the SPEED bench.

## 1. INTRODUCTION

Segmented telescopes offer a practical path towards dramatically enlarging telescope diameter from the ground (ELTs), or achieving optimal diameter in space. However, translating current technological advances in the domain of high-contrast imaging for monolithic apertures to the case of segmented apertures is far from trivial. Because of the primary mirror segmentation and the increased mechanical structures of the telescope such as the secondary mirror supports, the resulting pupil is geometrically complex. The pupil exhibits amplitude discontinuities created by the space between the segments and the presence of the secondary supports, and phase discontinuities resulting from imperfect alignment (phasing) between segments. These effects significantly limit high contrast imaging capabilities (speckles and diffraction), especially for exoplanet direct detection, which is a major driver for present and future observing programs. SPEED – the segmented pupil experiment for exoplanet detection – initiated in early 2013 at the Lagrange laboratory, aims at gearing up strategies and technologies for high-contrast instrumentation with segmented telescopes. It represents significant efforts in the vicinity of the ELT area, and future spatial mission by offering an ideal environment to progress in both phasing and high-contrast domains with complex/irregular apertures. It involves efforts in science-grounded instrument conception and design (optical, mechanical, and thermal engineering; development of novel techniques for tackling high-contrast imaging limitations such as coronagraphy, wavefront control and shaping), and will address several of the most critical issues barring the way for high-contrast imaging for the next generation of observatories.

* Email: patrice.martinez@oca.eu Tel: +33 (0)4 92 00 30 37

## 2. SCOPE OF THE EXPERIMENT

While the past decade has witnessed considerable technological advances in the domain of high dynamic range imaging for monolithic apertures, translating current technologies to the case of on-axis segmented apertures is far from trivial. The optical characteristics of a segmented telescope compose for a large part the error budget of static aberrations. This error sets the ultimate limit regardless the integration time. Tackling the various effects related to the segmentation in order to achieve a certain level of the image quality is fundamental [1, 2]. These effects can be separated into two categories: 1/ stochastic incoherent effects, where the value of characteristics changed randomly from segment to segment, which produce speckles; 2/ coherent effects, which produce a diffraction pattern with deterministic signature in the scientific image. Both wavefront control (correcting wavefront errors from imperfect optics, including segment phasing), and wavefront shaping (creating a dark zone in the PSF using deformable mirrors for amplitude and phase control) are a must with any segmented telescope.

For the stochastic effects (1/): the segments must be precisely phased by adequate wavefront control means. Recent laboratory and on-sky experiments have demonstrated precise segment phasing controlled at the telescope level [3], while phasing control at the level of an instrument has not yet being explored to our knowledge. If the telescope segment alignment error requirements expected by a high-contrast imaging instrument cannot be achieved by the telescope calibration (active optics of the telescope), phasing control by the instrument might be required. In addition to providing high-precision phasing feedback to the primary mirror, the use of a dedicated (segmented or continuous phase-sheet) deformable mirror (DM) for high precision phasing could be envisioned, being applicable for any segmented telescope. In particular, segment alignment defects in the primary mirror, consisting of a random combination of piston and tip-tilt errors on each segment, will produce discontinuities in the AO-corrected wavefront. While these residual phase errors will be too small to be detected by the telescope phasing system, their level, probably several tens on nanometers RMS, will be potentially prohibitive for exoplanet research.

For the deterministic effects (2/): optimal coronagraph designs and wavefront shaping must mitigate the bright PSF structures diffracted from the inter-segment spacing as well as for the structures diffracted from the secondary mirror support. Recent numerical research has generalized coronagraphic concepts to the case of segmented/irregular apertures [4, 5, 6, 7]. These results demonstrate that in the presence of a perfectly phased and non-aberrated segmented primary mirror, the diffracted structures from the pupil geometry (segment gaps, obstruction, support structures) can be reduced to very faint level. Further, inter-segment spacing (gaps), secondary support structures, can be treated as amplitude errors, and considered as such within wavefront shaping schemes [8]. Such generalization of amplitude error correction within an instrument has to be tackled and studied. The resulting trade-offs in coronagraph design, inner working angle, throughput, and wavefront control/shaping have yet to be fully explored.

On top of that segment-to-segment static optical amplitude errors must be accounted and corrected for (e.g., segment reflectivity dispersion, missing segments, etc). While it is commonly accepted that wavefront control/shaping on monolithic telescope apertures can be achieved using a linear approximation to model the relationship between actuator commands and electric field at the science image, the linear approximation is only valid for small errors and DM correction. By contrast, wavefront control and shaping solutions for segmented telescopes operate in a non-linear regime (segments exhibit misalignments with excursions initially much larger than the observing wavelength). As a consequence the usual linear approximations are not valid and thus non-linear wavefront control/shaping schemes might be considered and explored [8, 9]. The SPEED testbed aims investigating a practical solution for broadband coronagraphy on asymmetric, unfriendly apertures, with the developments of algorithms dedicated or optical approaches to minimize the segment effects and pupil discontinuity effects on the images and demonstrate the performance of the correction.

Over the past years, several high-contrast imaging benches, either dedicated to space-based applications [e.g., 9, 10, 11], or ground-based applications [e.g., 12, 13, 14], have been initiated and/or developed. In this context, the SPEED testbed has been though in a complementary manner to fit best the actual landscape without overlaps.

## 3. SYSTEM OVERVIEW

The section presents the description of the general layout of the SPEED facility with details of all systems and sub-systems that will be implemented and tested. Dedicated objectives will be discussed. The SPEED experiment can be described as the association of various independent instrumental modules (see Figure 1), with two optical paths, the visible path dedicated to segment phasing analysis, and the scientific arm in the near infrared operation in the H-band.

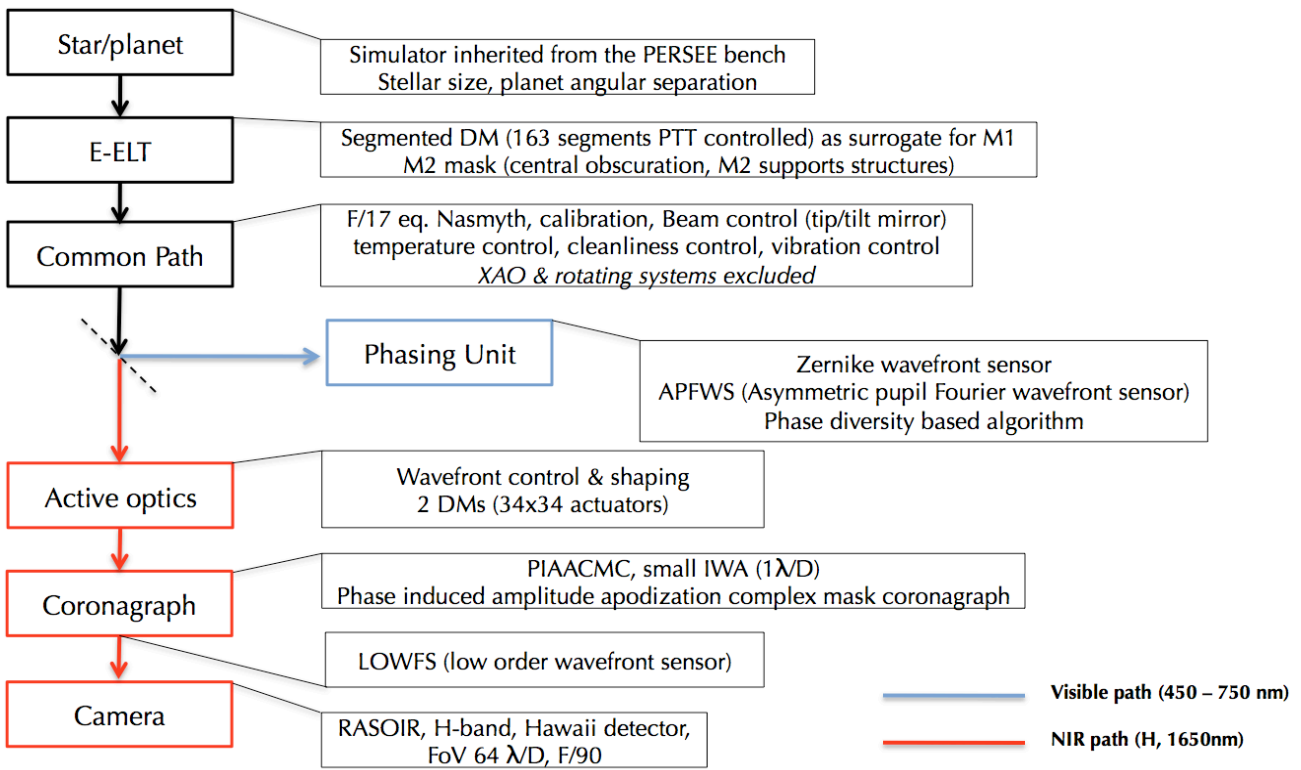

**Figure 1** SPEED general layout

### 3.1 Star/planet simulator

The SPEED bench will take avantage of an entirely designed and developed star and planet generator source module (called SPS, for "star and planet simulator") inherited from the PERSEE bench [15]. The SPS module allows to simulate a bright and unresolved star radiating from 0.6 to 3.3 μm, with an exoplanet signal orbiting around it with adjustable angular separation and flux ratio (varying from $10^{-4}$ to $10^{-6}$ in H-band). In practice, the "star" light is generated with a super-continuum source, while the "planet" light uses a Xenon lamp, and both are optically mixed (fibers are placed at the focus of two off-axis parabolic mirrors collimating the two exit beams toward a beam splitting plate). More information about the SPS module are largely available in the literature [15].

As ELTs will start resolving stars, stellar resolution up to 0.5 mas (utmost apparant stellar radii for nearby stars for a 39-m telescope, i.e., 0.03 λ/D at 1.6 μm) will be considered within the SPS module of SPEED. This aspect is extremely important for high-contrast imaging as small inner working angle (IWA) coronagraphs such as the PIAACMC are sensitive to the stellar angular size (coronagraph leakage due to stellar angular size is proportional to the square of stellar radii), and the leak is incoherent and will not interfere with speckles due to wavefront errors. Wavefront errors and coronagraphic leakage due to stellar angular size add incoherently in intensity, and cannot be corrected by speckle nulling techniques. Indeed, the PIAACMC will be conjointly optimized for complex aperture, IWA and stellar size.

### 3.2 Turbulence generator

The development of an extreme AO (XAO) system correcting for the short-lived and fast evolving atmospheric speckles, and hence the developpement of a turbulence generator per say, is out of the domain of interest of the SPEED facility. Within SPEED, we will assume that the atmospheric turbulence has been already corrected by an upfront XAO system, and concentrate on the quasi-static aberrations present inside the instrument. While the SPEED bench will mainly operate with the assumption that this virtual correction is perfect (no XAO residuals), atmospheric residuals left uncorrected by an XAO system will however be generated within the bench using exchangeable phase screens.

### 3.3 Segmented telescope simulator

The telescope simulator module consists in the association of a segmented mirror from IRIS AO vendor[1] [16] with 163 segments controlled in piston and tip/tilt (PTT489 DM) being used as a surrogate of the E-ELT (European-ELT) primary mirror, and an optical mask inserted into the beam to simulate the presence of the ELT secondary mirror (secondary mirror M2, including M1 central obscuration and M2 supports). The M2 mask consists in six radial arms regularly distributed (60 degrees between arms) and a large central obscuration (30%), though several configurations (i.e., several masks with various configurations) will be possible to be implemented and changed at any time. The segmented mirror (hereafter, SDM) is represented in Figure 2, where the inscribed pupil aperture is 7.7 mm diameter. Such a system will allow exploring several key aspects of the E-ELT architecture (missing segments, inter-segments spacing, island effect, etc.) regarding segment alignment (phasing), as well as the propagation of such errors and impact for high-contrast imaging.

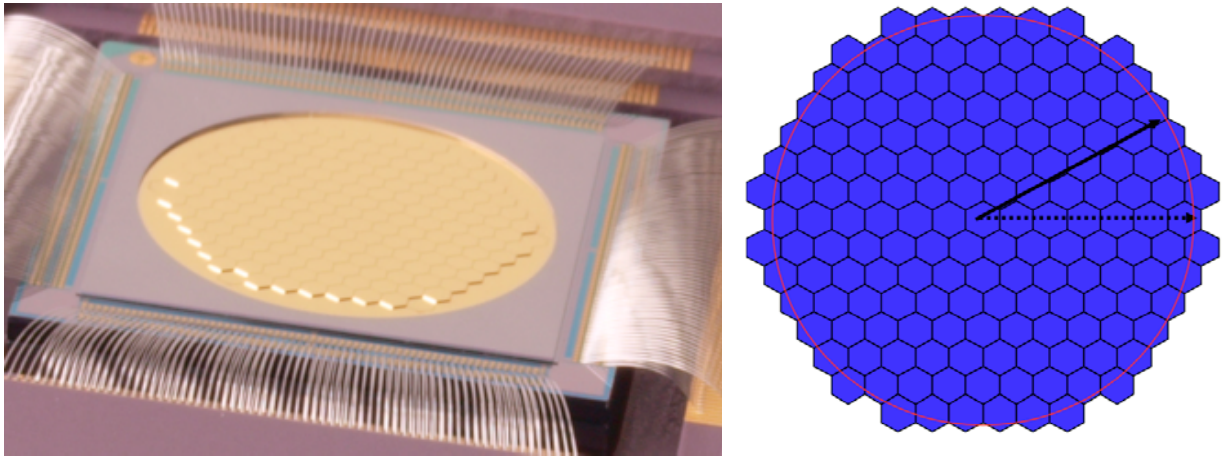

**Figure 2 Left: close view of a PTT489 mirror prototype, Right: schematic view of the 163 segments distribution over the inscribed aperture of 7.7mm (courtesy IRISAO vendor)**

### 3.4 Phasing unit

As part of the telescope simulator, a phasing unit system representative of what could be implemented on the active optics system of the future E-ELT is an attractive perspective and has been retained. In the framework of APE (the ESO Active Phasing Experiment [3]), four different phasing concepts have been compared under various conditions: ZEUS (Zernike Unit Sensor), SHAPS (Shack-Hartmann Phasing Sensor), PYPS (Pyramid Sensor), and DIPSI (Diffraction Intensity Phase Sensing Instrument). ZEUS [17] has been developed by LAM, IAC and ESO and is based on the Zernike Phase Contrast Sensor principle (which takes its origin in the Mach-Zehnder interferometer). The essential purpose of the APE experiment was to explore, integrate and validate non-adaptive segment phasing schemes and technologies at the telescope level for an ELT. Within SPEED, we propose to implement a Zernike phase contrast method, ZEUS-like system, while alternative phasing solutions, e.g., APFWS, the Asymmetric Pupil Fourier Wavefront Sensor [18], or phase diversity techniques will be implemented and compared.

The principle of the Zernike contrast phase sensor relies on the filtering of the low frequency components of the wavefront (filtered out) by a focal plane phase mask whereas the high-frequency part of the wavefront (wavefront discontinuities generated by the pupil aberrations) is leaving unchanged and finally appears as pupil intensity variations on a detector. These intensity variations, which are generated by the misalignment of the segments of the pupil, represent the signal of the system to be analyzed in order to retrieve the piston and the tip/tilt information. As a baseline two algorithm principles are being developed upon APE experience: the signal on the segment boundaries is extracted and can be either fitted to the theoretical signal, or divided into four different and identical areas where the intensity is integrated to retrieve the piston and tip-tilt components (see Figure 3 where the localization of the signal extraction is enlightened). The correction is then based for both approaches on the widely used method of interaction matrix construction and inversion matrix principle in an iterative way. Little iterations are generally necessary to reach optimal alignment upon initial amplitude error (the capture range of any phasing system is limited, though it can be increased with multi-wavelength or coherence based methods).

---

[1] http://www.irisao.com

While the current specification for an ELT segment phasing can reasonably be considered as achievable by such means, reaching the high-precision phasing requirements of a high contrast instrument is uncertain. Although it would probably be possible to achieve much higher correction with a closed-loop operation of the M1 primary mirror, it might drive complexity for the telescope. Phasing control by the instrument might then be considered, though phasing control at the level of the instrument have not yet being explored, and will therefore be tackled within the SPEED testbed. On top of that, phasing defects residual from the telescope control will produce discontinuities in the AO-corrected wavefront error of several tens of nanometer rms. This adds a layer of complexity to the wavefront control and shaping problem for high contrast imaging instrument. Fine segment phasing residual correction can in principle be achieved through standard continuous phase-sheet DM in the instrument XAO closed-loop, but this would place a substantial load on the wavefront control system by using part of the DM stroke for phasing. A fully dedicated DM for high precision phasing together with adapted control algorithm is one way to solve the problem. As phasing defects residual are mixed up with AO-corrected wavefront residuals, a continuous phase-sheet DM might be well suited to do the job. Another way to solve the problem is to develop algorithms to extract segment misalignment residual error information from post AO-wavefront measurements that can be fed back to the correcting system, either directly to the telescope primary mirror, or to the correcting system that would be integrated in the instrument (DM). These problematic will be treated within the SPEED bench. Alternative solutions with new features might be considered at a later stage using post-coronagraphic image analysis with dedicated wavefront sensor.

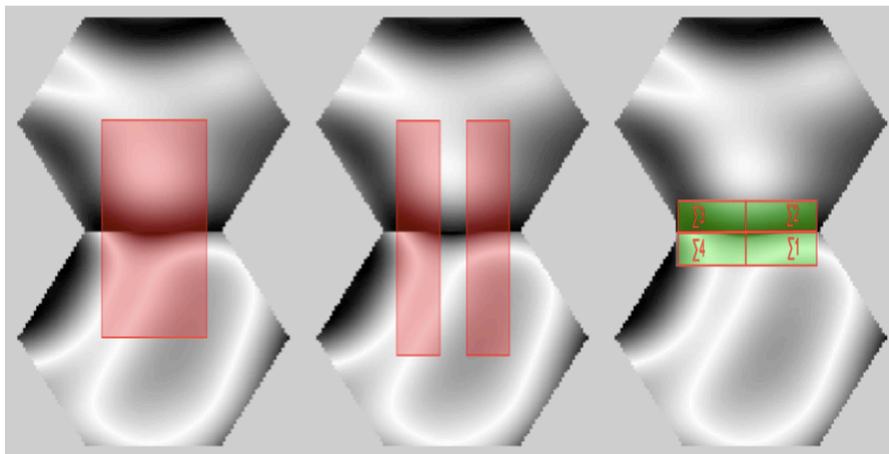

**Figure 3 Extraction of the signal for two different algorithms within the Zernike contrast phase sensor. Left and middle: 1$^{st}$ algorithm area of signal extraction for piston (left) and tip-tilt (middle, carried out through differential piston measurement) estimation through fitting on theoretical signal originating from the Mach-Zehnder theoretical signal, Right: 2$^{nd}$ algorithm based on the signal integral estimation over four separated but identical areas at the segment boundaries.**

### 3.5 Wavefront control and shaping

Two sequential DMs separated by free-space propagation can be efficiently used to correct for both phase and amplitude errors on a symmetrical field of view for relatively wide spectral bandwidth, while two DMs in non-conjugate planes close to the pupil can also remap the pupil discontinuities [8]. With the same philosophy as for the PIAACMC (see next sub-section), the upfront correction of aperture irregularities by optical remapping in the geometric and thus achromatic regime is feasible in order to feed it into a coronagraphic stage. ACAD, the Active Compensation of Aperture Discontinuities [8] has been recently proposed, to derive mirror shapes suitable to remove the structures introduced by spiders, gaps without loosing photons. Because the required mirror deformations are small and are of the order of a few microns, DMs could be used for this purpose to reach high contrast level (e.g., $10^{-8}$) for ELTs. DM1 and DM2 of SPEED from Boston Micromachines vendor[2] can be used in an ACAD configuration optimized together with PIAACMC to deliver high contrast level with complex aperture. The aperture discontinuities correction will likely both required ACAD and PIAACMC optimization to obtain high contrast level, which would otherwise place a too high load on DM1 and DM2. DM1 and DM2 are Kilo-C-DM with 34x34 actuators across the circular pupil (952 actuators in total) with 1.5 μm stroke (300 μm pitch).

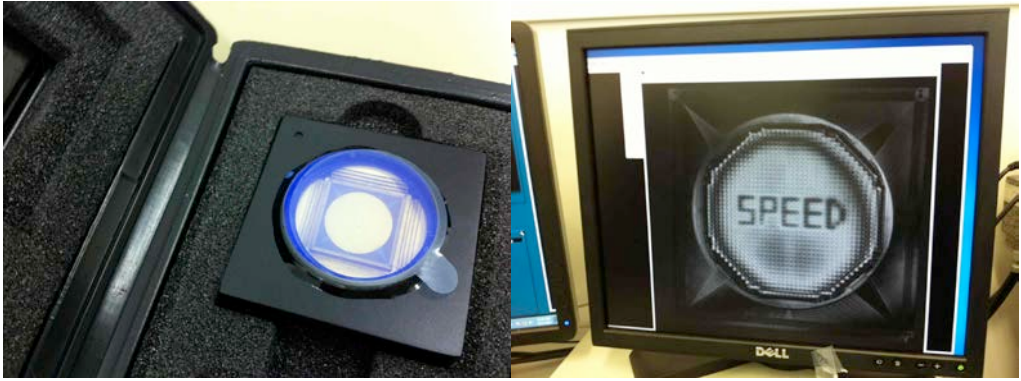

**Figure 4 Left: picture of one of the two continuous sheet deformable mirror (Kilo-C-DM) from Boston Micromachines vendor, Right: image taken with a Zygo interferometer.**

### 3.6 Coronagraphy

Several small inner-working angle class coronagraphs are currently subject to vigorous R&D to gear them up to high technological maturity. In particular, the PIAA (Phase Induced Amplitude Apodization, [19]) uses beam remapping for lossless apodization and can be combined with opaque masks (PIAAC and PIAALC) or partially transmissive phase-shifting (complex) masks (PIAACMC [7]). It theoretically offers complete starlight extinction, with high throughput and sub-$\lambda/D$ inner working angle, regardless of the aperture shape. The PIAA offers nearly 100% throughput and approaches the fundamental coronagraph performance limits [20]. The technological maturity of the PIAA benefits from 10 years R&D activities, from scratch towards on sky demonstration, and actual raw contrast reaches level up to $10^{-8}$ at 2 $\lambda/D$ (on going improvements are foreseen to obtain $10^{-9}$) in monochromatic light (NASA Ames coronagraphic testbed), and expected to deliver $10^{-6}$ in H-band at 1 $\lambda/D$ at the Subaru telescope. The goal within SPEED is to develop a PIAACMC that can cope with $1\lambda/D$ IWA (inner working angle), stellar angular size up to 0.5 mas, and correct for all or part of the telescope pupil discontinuities (secondary support structures), with $10^{-7}$ raw contrast at IWA. As the PIAACMC development will likely require intensive efforts in term of optimization process, manufacturing and testing, a step 0 coronagraph (conventional pupil apodization [21]) will be considered first on the SPEED testbed to open the path towards PIAACMC at a later stage during the project.

### 3.7 Low order wavefront sensor

A low order wavefront sensor is critical to ensure that starlight remains centered at the coronagraph plane. Because the PIAACMC is apart of a family of coronagraph that is optimized for the detection of companions at very low angular separations, it is highly sensitive to low-order aberrations, especially tip/tilt errors (pointing errors). SPEED therefore includes a robust and efficient wavefront sensor to measure tip-tilt as well as defocus, namely the Coronagraphic Low-Order Wavefront Sensor (CLOWFS) currently in use in the SCExAO instrument at the Subaru telescope [22, 23]. Since the PIAACMC is a phase mask based coronagraph, the light used for the analysis is taken at the Lyot-stop plane, i.e., reflected towards an optics that refocuses the light onto the LOWFS camera. A dedicated tip-tilt mirror upstream in the optical path (see Figure 1) will be used to correct for the tip-tilt estimation from the CLOWFS, though tip-tilt correction could be carried out alternatively by DM1, DM2, or SDM. As the PIAACMC has to cope with stellar angular size up to 0.5 mas (0.06 $\lambda/D$ at 1.6 μm), residual jitter (tip-tilt) must be no larger than the stellar angular size in the error budget, hence no larger than few $10^{-2}$ λ/D at 1.6 μm, and on top of that to achieve a $10^{-7}$ PSF raw contrast at such small IWA with the PIAACMC, quasi-static pointing should likely be accurate to about $10^{-3}$ λ/D at 1.6 μm. Such tiny levels have already been demonstrated at a $10^{-2}$ level at the Paris Observatory [24] and $10^{-3}$ level at the Subaru telescope [22, 23].

---

[2] http://www.bostonmicromachines.com

### 3.8 Scientific cameras

The infrared camera is working at 1.65 μm with an internal H-band filter. Its read-out-noise is 12 e- rms/pixel and quantum efficiency > 60%. The detector pixels (18.5 μm) are read in double correlated sampling mode, and is an 1kx1k Hawai array (engineering grade) from which we select only a quadrant of 512 pixels, enough considering the field of view of 64 λ/D. The camera has been kindly lending by the European Southern Observatory[3].
The optical camera for the phasing unit is an Apogee camera with 1024 x 1024 pixels (13 μm pixel) with 2.2 e-rms/pixel read-out-noise, and 92% quantum efficiency.

## 4. INSTRUMENTAL AND CONTRAST DESIGN

### 4.1 Rationale

In context of a high-contrast imaging instrument, the relationship between scientific requirements (performance requirements) and instrumental requirements is not trivial and generally quite complicated. Current exoplanet imager systems are rarely limited by residual atmospheric speckle noise, but rather by quasi-static speckles, evolving on larger timescales and originating from internal aberrations sources [e.g., 25, 26, 27]. Contrast is usually determined by various factors including cross talk effects, and several guiding rules are generally considered:
- All major and independent contrast terms must be set to/or below the order of magnitude of the final target contrast requirement. As a matter of fact this is not sufficient when one consider cross-terms. Cross-terms contribution must be set comparable as the target contrast that would degrade contrast otherwise.
- All error sources must be minimized from chromaticity point of view (e.g., refractive optics, chromatic beam-walk, Fresnel propagation, etc).
- Optics must be accurately specified to reach the best quality considering low, middle, and high frequency contents. Various terms potentially affecting the target contrast (Fresnel effect, non-common path aberrations, etc) derive from optical quality of individual surfaces.

In this context, an additional and specific care must be considered regarding Fresnel and Talbot effects, which are fundamentally uncorrectable when the optical design is frozen, and are design/system dependent. Fresnel propagation terms and Talbot effect are important class of perturbation affecting a high contrast imager (a pure phase aberration on an optical surface mixes between phase and amplitude aberrations as light propagates). In a collimated beam, this oscillation effect occurs over a distance called the Talbot distance being proportional to the square of the aberration spatial period and inversely proportional to the wavelength of light (hence the Talbot distance is chromatic). This effect is particularly important as phase aberrations if correctly sensed can be corrected with a single deformable mirror, but amplitude aberrations cannot. To correct for amplitude aberrations in addition to phase aberrations one must measure amplitude contribution and operate two sequential deformable mirrors (the second takes advantage of the Talbot effect, correcting for phase aberrations that were originally amplitude aberrations but have oscillated into phase after a given optical distance of light propagation). With these considerations, two important aspects must be treated:
- As these effects are inherently chromatic and design dependent, they must be minimized by optical design.
- Optical phase and amplitude errors can be corrected using two-sequential DMs assuming well mastering of the Fresnel/Talbot effects, in particular in the positioning of the second mirror with respect to the first one is fundamental and dependent on the aberration spatial frequency needed to be corrected.

The SPEED testbed optical development will hence rely on a three parallel approaches carried out in an iterative process: ray tracing analysis, analytical sensitivity analysis, and end-to-end Fresnel extensive simulation work, in order to optimally converge to an optical and mechanical design that can cope with the initial objectives of the project. The opto-mechanical design presented in the next sub-sections is then premature and will be subject to modifications in the next months, and must be understood as a first step towards a final version planned for fall 2014 or early 2015.

---

[3] http://www.eso.org

**4.2 Pre-optical design**

The SPEED testbed will be installed on a 1.5 x 2.4 m table with air suspension, and reflective optics have been adopted for the whole design, except for the phasing unit arm. Few sub-systems that have been described in the previous section put some constraints on the design (DM1, DM2, SDM, NIR camera, and the star and planet generator). The testbed will be located in a clean room with airflow during the integration and alignment phase, while ultimately operated completely covered with protection panels forming a nearly close box with the objective of minimizing the internal turbulence, and optimizing stability. Temperature, humidity, pressure, and stability will be monitored using probes, accelerometers, lasers and laser beam positioning sensor, etc., and the bench environment is under intensive characterization. The stability of the whole system is apart of the major concerns at the early stage of the bench development, and will be characterized (room, table, sub-systems), monitored, pre-compensated by designed as far as possible, and efforts will focus on developing actively controlled solutions to ensure the required level of stability for reaching the objectives of the project.

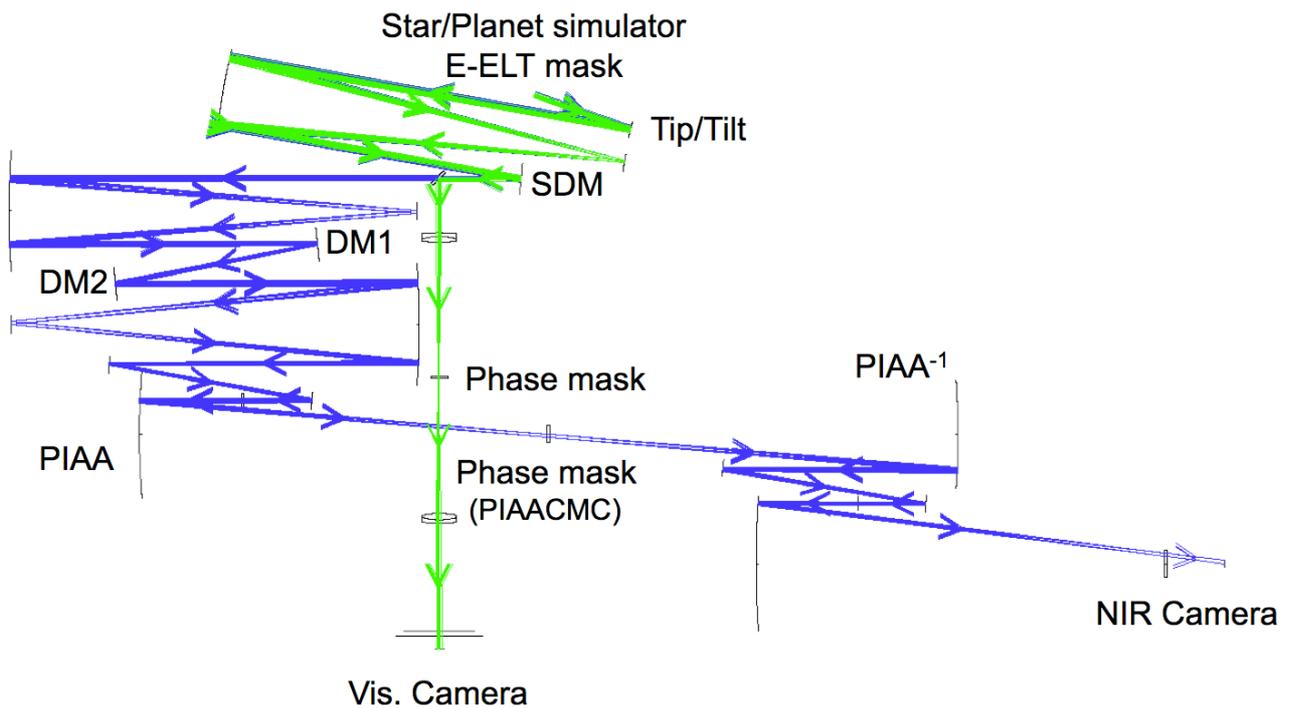

**Figure 5 Optical design of the SPEED bench using ray tracing software.**

The preliminary optical design is presented in Figure 5. The entrance F-number is F/17 as for the E-ELT Nasmyth port, and the pupil is defined in the plan where the SDM is located with 7.7 mm diameter. DM1 is in a conjugated pupil plane, while DM2 is out of pupil plane. The PIAACMC will be entirely developed with reflective optics and the F-number at the focal plane mask is F/60 and F/90 at the entrance of the NIR camera. The field of view at the NIR detector plan is 64 x 64 $\lambda/D$ with a sampling of 8 pixel per $\lambda/D$. The F-number at the visible focus of the phasing unit is F/15. All movable components (filters, neutral densities, phase masks) will be inserted on motorized translation/rotation stages to guaranty for positioning accuracy, stability, and reproducibility. In addition, an integrating sphere will be located at the entrance of the bench for calibration purpose. A 3D view of the SPEED testbed is presented in Figure 6.

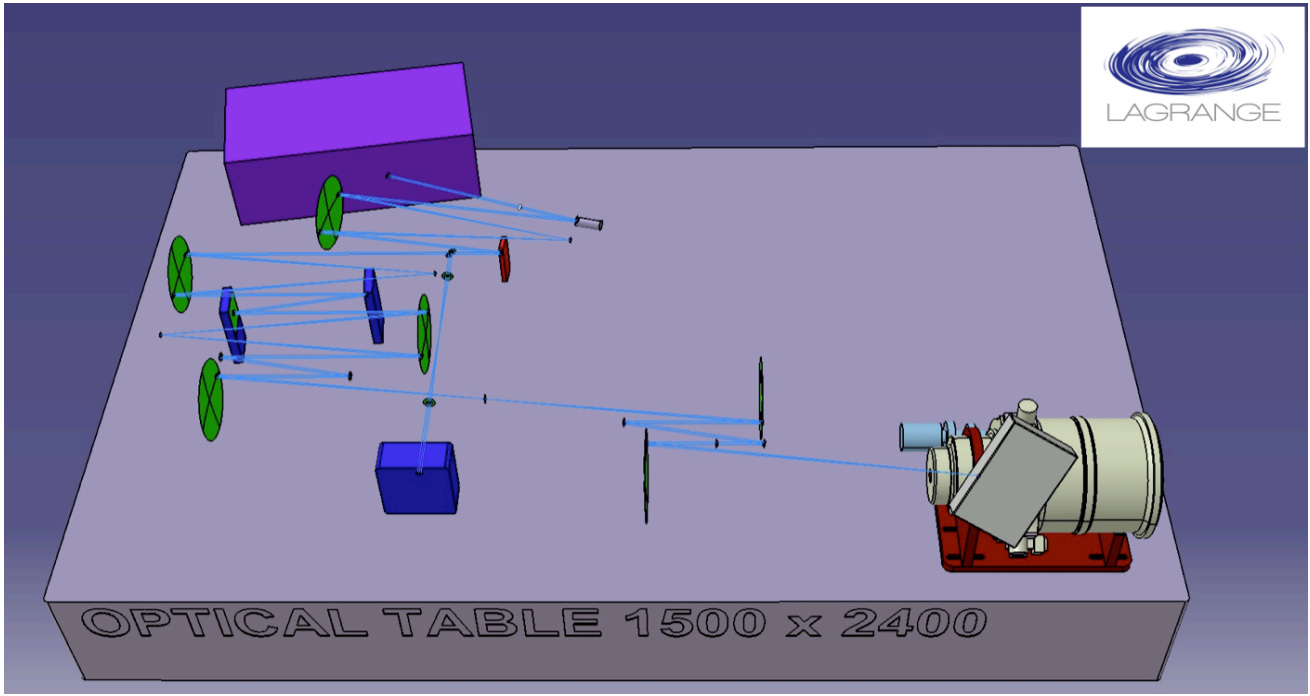

**Figure 6 Overview of the SPEED bench represented in a 3D optical/mechanical view.**

## 5. CONCLUSION

The SPEED bench is under extensive development at the Lagrange laboratory, and will be unique in Europe for tackling high-contrast imaging with segmented and complex/irregular telescope aperture. While the SPEED optical design finalization is under way, pending to extensive numerical simulation efforts around the problematic of Fresnel/Talbot effect, the integration is planned to start in early 2015.

## 6. ACKNOWLEDGEMENTS

The activity outlined in this paper is partially funded by the Région Alpes Côte d'Azur, by the French government, and by the European Union as part of the FEDER program (contract number 1 083-40 624). The project also benefits from funding support of the Nice Sophia-Antipolis University, Côte d'Azur Observatory, and Lagrange laboratory. The authors warmly thank Pierre Baudoz and Raphaël Galicher from LESIA, C. Vérinaud and Alexis Carlotti from IPAG, for fruitful discussions about Fresnel/Talbot effects. We are grateful to ESO for the loan of the NIR camera, and warmly acknowledge Sebastien Tordo, Markus Kasper, and Norbet Hubin.